\documentclass[prd,aps]{revtex4}
\usepackage{amsmath,amssymb,amsfonts,dcolumn,color,graphicx,graphics,latexsym,placeins,epsfig}
\usepackage[compatibility=false]{caption}
\usepackage{subcaption}
\usepackage{hyperref}

\usepackage{natbib}
\usepackage{float}
\usepackage{booktabs}
\usepackage{array}
\newcommand{\be}{\begin{equation}}
\newcommand{\ee}{\end{equation}}
\newcommand{\ba}{\begin{eqnarray}}
\newcommand{\ea}{\end{eqnarray}}

\begin{document}

%\preprint{LIGO-P1800035}

\title{\Large \bf Quasi-normal modes in a symmetric triangular barrier}
\author{Poulami Dutta Roy${}^{1}$, Jagannath Das${}^{1}$ and
Sayan Kar ${}^{1,2}$}
\email{poulamiphysics@iitkgp.ac.in, jagannath.das0111@gmail.com, sayan@iitkgp.ac.in}
\affiliation{${}^1$ Department of Physics \\ Indian Institute of Technology Kharagpur, 721 302, India.}
\affiliation{${}^{2}$Centre for Theoretical Studies \\ Indian Institute of Technology Kharagpur, 721 302, India. }

\begin{abstract}
\noindent Quasi-normal modes (QNMs) of the massless scalar wave in $1+1$ dimensions are obtained for a symmetric, finite, triangular barrier potential. This  problem is exactly solvable, with Airy functions involved in the solutions. Before obtaining the QNMs, we demonstrate how such a triangular barrier may arise in the context of scalar wave propagation in a tailor-made wormhole geometry. Thereafter, the Ferrari-Mashhoon idea 
is used to
show how bound states in a well potential may be used to find the QNMs in a corresponding barrier potential. The bound state condition in the exactly solvable triangular well and  the transformed condition for finding the QNMs are written down.
Real bound state energies and complex QNMs are found by solving the respective transcendental equations.
Numerical integration of the wave equation yields the time domain profiles for scalar waves propagating in this wormhole geometry which illustrate the quasinormal ringing.  
Estimates relating the size of the wormhole throat (in units of solar mass) with the
QNM frequencies are stated and discussed. 
Finally, we show how the effective potential and the QNMs for scalar perturbations
of the Ellis--Bronnikov wormhole spacetime can be reasonably well--approximated using a properly 
parametrised triangular barrier.

\end{abstract}

%{Classical general relativity, Exact solutions}

\maketitle

\section{\bf INTRODUCTION} 

\noindent An oft-quoted example of the quasinormal mode is the ringing sound
we hear when we strike a metal glass with a metallic rod. The ringing decays
in time--thus the mode is {\em quasi}-normal and not a usual normal mode.
The decay of the ringing obviously means that energy is lost in the environment
--the system is open and hence dissipative. In many ways, such systems are
interesting and also realistic. 

\noindent The earliest mention and discussion on such modes appear in the
paper in 1900 by Horace Lamb \cite{lamb} who calls them {\em peculiar vibrations}, while 
analysing the wave system arising in the free vibrations of a nucleus in an 
extended medium. Much of the early work on these modes (and also quite a bit
of the later work as well) are thus in nuclear physics.

\noindent  In the late fifties, the main governing equations for gravitational perturbations of black holes (eg. the Schwarzschild) were derived in the context of the stability of the
Schwarzschild spacetime in the
seminal paper by Regge and Wheeler \cite{regge}. Much later, Zerilli \cite{zerilli} obtained the effective potential for even-parity perturbations.
However, it was in 1970, in an article published in the journal {\em Nature} \cite{Vishveshwara}, Vishveshwara
first noted the quasi-normal mode in 
gravitational physics, while working on perturbations of
black holes \cite{Vishveshwara1}. Following soon, came the work of Chandrasekhar and
Detweiler in 1975 \cite{Chandrasekhar}, wherein, among other issues concerning the Schwarzschild
black hole, the exact quasinormal modes in a toy model of
the rectangular barrier potential was discussed. Several other potentials
also exhibit exact quasinormal modes (eg. Poschl-Teller \cite{PT},\cite{PT1}), which have been
found and analysed in detail, over the years.  A fairly comprehensive list of
such potentials and their quasi-normal spectra can be found in 
\cite{visser}.

\noindent In the current wake of detection of gravitational waves \cite{Ligo1} the study of quasi-normal modes has received renewed interest. 
Such modes arise
in the context of black hole stability and are discussed extensively in black hole
perturbation theory. Since there is loss of energy involved, the quasi-normal frequency must be complex with the real part giving the actual energy carried by the wave and the imaginary part giving an idea about the damping time.  
One must have the right sign for the imaginary part in order to have a decay in the time domain profile. 
In astrophysics, the QNMs are important because they depend only on the source/system parameters (eg. mass, angular momentum etc.) and not on the cause producing them.  This can help us in obtaining information about the parameters of a black hole or even a wormhole \cite{aneesh}. Through the nature of the solutions for the scalar or tensor
QNMs, one can differentiate various theories of gravity \cite{f(R)}. %the equation of state of a neutron star.\\

\noindent The central problem lies in calculating the modes. In most cases, the wave equation is not exactly solvable 
due to the complexity of the effective potential.
Numerical methods are the only way out. Though various semi-analytical techniques, like the WKB method \cite{WKB} are there to provide a physical intuition on the behavior of these modes, to get precise values, one has to 
use different numerical techniques. The handful of scenarios where exact solutions are possible are
limited and therefore important. Such exact solutions, if not of any real value (in the sense of observations), 
do have great utility in a pedagogic sense, to say the very least. 
As mentioned before, apart from the square barrier, there are several other
potentials for which exact expressions for the modes are available \cite{visser}. In our work here, 
we study another simple example--the case of a symmetric, finite, triangular barrier.

\noindent The triangular barrier potential and its inverse (the triangular well) are exactly solvable potentials whose solutions are found in terms of the Airy functions \cite{ghatak}. The knowledge of the bound states of the
potential (in terms of Airy functions) shows the way towards finding the QNMs, using an interesting 
correspondence first found by Ferrari and Mashhoon \cite{mashhoon1,mashhoon2,ferrari,blome}. The Ferrari-Mashhoon method  shows how a given 
barrier potential, via a
change of variables and parameters, can be inverted into a well potential, 
in which bound states 
can be found. Using the inverse transformation, one can extract the QNMs too, including the spatial wave functions. 
The QNMs which are obtained from inverse transformation of the bound state eigenvalues are termed as the proper quasi-normal modes. However, the QNMs can also be found by directly using the appropriate
boundary conditions and without referring to the corresponding bound-state problem in an
inverted (well) potential. Despite being an interesting idea, the Ferrari-Mashhoon correspondence has limited
applicability in finding QNMs, largely because of its reliance on the knowledge of the solution of the associated bound state problem.

\noindent Our work here is organised as follows. In Section II we briefly introduce the potential well and also the barrier on which our work is based. Also, in this section we construct a wormhole solution \cite{morris} whose effective potential for scalar wave propagation is a triangular barrier. In Section III, we discuss the idea of quasi-normal modes and explain the Ferrari-Mashhoon
correspondence. Section IV deals with the bound states of the symmetric, finite triangular well and the corresponding energy eigenvalues. 
Section V is devoted to QNMs of the triangular barrier. Here we first find the QNMs using the Ferrari-Mashhoon idea and then mention how the QNMs can also be obtained independently. 
Thereafter, we numerically solve the wave equation and 
obtain the time-domain profiles at a fixed value of the spatial
coordinate. In order to get some idea about numbers, we calculate the frequency and corresponding damping time for different throat radii of the wormhole. Finally, we show how the effective potential of the well-known
Ellis--Bronnikov spacetime \cite{ellis} can be well-approximated by a triangular
barrier.
Section VI is a summary with some comments on the results obtained and 
possible implications and extensions. A short conclusion appears in Section VII.

\section{TRIANGULAR WELL AND BARRIER POTENTIALS}
\noindent The symmetric, finite triangular potential well is given as,
\begin{gather}
   \mathrm{V(x) }=\begin{cases} \mathrm{\frac{V_{0} |x|}{a} - V_{0}} & ; \,\, \mathrm{-a\leq x \leq a}\\
         0         & ; \,\,\mathrm{ x > a, x < -a  \,} ,
        \end{cases}
\end{gather}
\noindent where $2 a$ is width of the well and $V_{0}$ is the depth of the well. 
In order to get the corresponding potential barrier we invert the above potential,
\begin{gather}
   \mathrm{V(x) }=\begin{cases}\mathrm{V_{0} - \frac{ V_{0} |x|}{a}  } & ; \,\,\mathrm{ -a\leq x \leq a}\\
         0         & ; \,\, \mathrm{x > a, x < -a  \,} .
        \end{cases}
\end{gather}
Figures 1(a) and 1(b) show the triangular well and the triangular barrier, respectively.

\noindent  Let us now see where, and in which context, a triangular barrier potential may arise. It is a known fact that the effective potentials for scalar wave propagation
 in wormhole geometries \cite{morris} have single or double barrier features. For example, in Ellis-Bronnikov geometry \cite{ellis} 
 the effective potential is a single barrier. Can we have a wormhole geometry for which the
 effective potential is a triangular barrier? To address and answer this question, we do 
 some reverse engineering. We write down the scalar wave equation in a spacetime with one unknown function $r(x)$. Thereafter we obtain the $r(x)$ for which we have a wormhole.
 
\noindent  To begin, let us assume the line element 
 to be of the form:
\begin{align}
    \mathrm{ ds^{2} = -dt^{2}+dx^{2}+r^{2}(x)(d\theta^{2}+sin^{2}\theta d\phi^{2}),}
 \end{align}
 where the coordinate $x$ ($-\infty \leq x\leq \infty$) is like the `tortoise' coordinate commonly used in black hole physics and the radial coordinate $r$ is written in terms of $x$ through the relation
 \begin{align}
    \mathrm{ dx^{2}= g_{rr} {dr^{2}}\,} .
 \end{align}
 The only unknown function in the line element is $r(x)$ which, for a wormhole, is
 symmetric, $r(x)= r(-x)$. Also, $r(x=0)$ is finite, positive and non-zero. In addition, for asymptotic flatness,
 we usually assume that $r\rightarrow \pm x$ as $x$ becomes large. However, in our 
 example below, we will see that $r(x)$ will tend to $\pm x$ at a finite $x$, exactly where
 the triangular barrier will end and the potential will become zero. In other words,
 we have a wormhole `sandwiched'
 between two flat regions.
 
 \vspace{0.15in}
 \begin{figure}[H]
      \centering
 \begin{subfigure}[t]{0.45\textwidth}
 	\centering
 	\includegraphics[width=\textwidth]{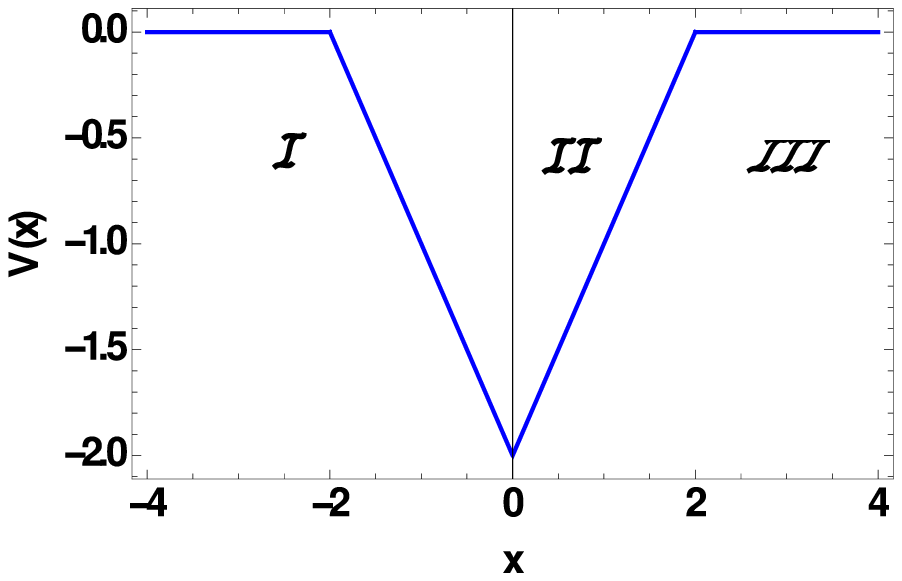}
 	\caption{Potential well}
 	\label{fig:fig(a)}
 \end{subfigure}
 \hspace{0.4in}
 \begin{subfigure}[t]{0.45\textwidth}
 	\centering
 	\includegraphics[width=\textwidth]{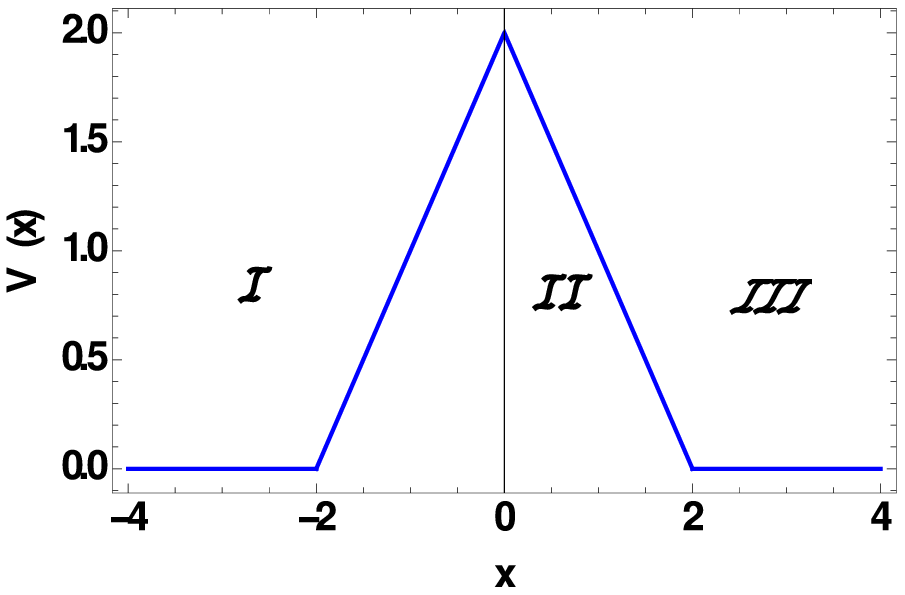}
 	\caption{Potential barrier }
 	\label{fig:fig(b)}
 \end{subfigure}
 \caption{$V_{0} = a = 2$}
     \label{fig:fig1}
\end{figure}

 \noindent If we intend to study the propagation of a massless scalar wave in such a 
 spacetime we need to solve the Klein-Gordon equation, given as:
\begin{align}
    \Box \Phi = 0 \, .
\end{align}
After separation of variables using the ansatz
\begin{align}
    \mathrm{\Phi(x,t,\theta,\phi) = u(x)\, e^{-i \omega t}\, Y(\theta \,,\phi)}
\end{align}
and writing the radial equation, we get
\begin{align}
   \mathrm{ \frac{d ^{2} u}{d x^{2}} + \frac{2 r'}{r} \frac{d u}{d x}+\Big(\omega^2 - \frac{m(m+1)}{r^2}\Big) u = 0 \, },
\end{align}
where the prime denotes a derivative w.r.t. $x$. $m$ is related to the angular momentum.
To find the effective potential, we rewrite equation (7) in the form of the time-independent
Schr\"{o}dinger equation,
by eliminating the first order derivative term with a change of the
dependent variable. Thus, assuming $u(x) = A(x) B(x)$ and considering the differential equation for $A(x)$, we note that for the first derivative in $A(x)$ term to  vanish, we need $B(x)$ to satisfy
\begin{align}
   \mathrm{ \frac{B'}{B} = - \frac{r'}{r}} .
\end{align}
Substituting this in equation $(7)$ we find the equation for $A(x)$ to be
\begin{align}
   \mathrm{ \frac{A''}{A} + \Big(\omega^{2} - \frac{m(m+1)}{r^{2}} - \frac{r''}{r}\Big) = 0 \, }.
\end{align}
Hence the effective potential is given as,
\begin{align}
   \mathrm{ V(x) = \frac{m(m+1)}{r(x)^{2}} + \frac{r''(x)}{r(x)}\,} .
\end{align}
If we want the effective potential for the
$m=0$ mode to be a symmetric triangular barrier as shown in Figure $1(b)$ then,
\begin{align}
   \mathrm{\frac{r''}{r}} =\begin{cases}\mathrm{V_{0} - \frac{V_{0} |x|}{a}  }& \mathrm{,-a\leq x \leq a}\\
    0 & \mathrm{ ,x > a, x<-a\,} .
    \end{cases}
\end{align}
On solving the above equation we get $r(x)$ given as,
\begin{align}
   \mathrm{ r(x)} =\begin{cases} \mathrm{-x} & \mathrm{,x<-a}\\
  \mathrm{C\, Ai\Big[\frac{V_{0}\,+\frac{V_{0} x}{a}}{(\frac{V_{0} }{a})^{\frac{2}{3}}}\Big] + D\,  Bi\Big[\frac{V_{0}\,+\frac{V_{0} x}{a}}{(\frac{V_{0} }{a})^{\frac{2}{3}}}\Big]} & \mathrm{,-a\leq x \leq 0}\\
     \\
   \mathrm{ C\, Ai\Big[\frac{V_{0}\,-\frac{V_{0} x}{a}}{(\frac{V_{0} }{a})^{\frac{2}{3}}}\Big] + D\,  Bi\Big[\frac{V_{0}\,-\frac{V_{0} x}{a}}{(\frac{V_{0} }{a})^{\frac{2}{3}}}\Big]} & \mathrm{ ,0 \leq x \leq a}\\
   \mathrm{ x }& \mathrm{,x>a}
  \end{cases}
\end{align}
where Ai and Bi are the Airy functions of first and second kind respectively and $C$ and $D$ are constants. For the regions $x>a$ and $x<-a$ the spacetime is flat 
and $V(x)=0$. The metric in these two regions is Minkowskian. One may note that the differential equation for region $0\leq x \leq a$ will have a general solution with the argument of the Airy function being $\frac{V_{0}-\frac{V_{0} x}{a}}{(\frac{-V_{0} }{a})^\frac{2}{3}}$. Since we require $r(x)$ to be a continuous function over the entire region, we take only one root of $(-1)^\frac{1}{3}$ i.e. -1, which, when squared becomes 1. The other two roots of $(-1)^\frac{1}{3}$, which we do not use, correspond to Airy functions having factors $e^{\frac{\pm 2i \pi }{3}}$ in the argument.\\
We need to ensure that the function $r(x)$ thus constructed, 
is continuous and so is its first derivative across the boundary. This requirement will
lead to the following two matching conditions:
\begin{gather}
  \mathrm{  C\, Ai [0]+ D\, Bi[0] = a}\\
   \mathrm{ C\, Ai'[0] + D\, Bi'[0] = -\Big(\frac{V_{0}}{a}\Big)^{\frac{-1}{3}}\,} .
\end{gather}
Taking the ratio of the above equations we obtain,
\begin{align}
 \mathrm{\frac{C}{D} =-\frac{z Bi'[0] + Bi[0]}{z Ai'[0] + Ai[0]}}
\end{align}
where $z=(V_{0} a^{2})^{\frac{1}{3}}$.

\noindent Since we wish to construct a wormhole geometry, $r(x=0)$ cannot be zero. Hence we must consider the even set of solutions for which $r'(x=0)=0$. Imposing this condition we find,
\begin{align}
\mathrm{ \frac{C}{D} = -\frac{Bi'[z]}{Ai'[z]}\,} .
\end{align}
Thereafter, combining equations $(15)$ and $(16)$, we arrive at a transcendental equation of the form-\\
\begin{align}
   \mathrm{ \frac{Bi'[z]}{Ai'[z]} = \frac{z Bi'[0] + Bi[0]}{z Ai'[0] + Ai[0]}}.
\end{align}
Solving the above equation, gives the value of $z=1.58377$.\\
This value imposes a condition on $V_{0}$ and $a$ given as,\\
\begin{align}
   \mathrm{ z = (V_{0} a^{2})^{\frac{1}{3}} = 1.58377}
\end{align}
which, in turn implies that the $m=0$ mode effective potential of the scalar wave
(spatial part) is a symmetric triangular barrier.\\
The constants $C$ and $D$ have the dimension of length due to the overall factor `$a$' present in their expressions. Solving equations $(13)$ and $(14)$ we obtain,
\begin{gather}
    \mathrm{C\,= a\,{\frac{Bi'[0]+\frac{1}{z}Bi[0]}{Ai[0]Bi'[0]-Ai'[0]Bi[0]}}\,= a{\frac{0.44 + \frac{1}{z}0.61}{0.318}}}\\
   \mathrm{ D\,= a\,{\frac{Ai'[0]+\frac{1}{z}Ai[0]}{Ai'[0]Bi[0]-Ai[0]Bi'[0]}}\,= a{\frac{-0.26 + \frac{1}{z}0.35}{-0.318}}\,} .
\end{gather}
Finally, we arrive at the expression of $r(x)$ given as,
\begin{align}
  \mathrm{r(x)} =\begin{cases} \mathrm{-x} & \mathrm{,x<-a}\\
 \mathrm{a{\frac{0.44 + \frac{1}{z}0.61}{0.318}}\, Ai\Big[\frac{V_{0}\,+\frac{V_{0} x}{a}}{(\frac{V_{0} }{a})^{\frac{2}{3}}}\Big] +  a{\frac{-0.26 + \frac{1}{z}0.35}{-0.318}}\,  Bi\Big[\frac{V_{0}\,+\frac{V_{0} x}{a}}{(\frac{V_{0} }{a})^{\frac{2}{3}}}\Big]} & \mathrm{,-a\leq x \leq 0}\\
     \\
   \mathrm{ a{\frac{0.44 + \frac{1}{z}0.61}{0.318}}\, Ai\Big[\frac{V_{0}\,-\frac{V_{0} x}{a}}{(\frac{V_{0} }{a})^{\frac{2}{3}}}\Big] + a{\frac{-0.26 + \frac{1}{z}0.35}{-0.318}}\,  Bi\Big[\frac{V_{0}\,-\frac{V_{0} x}{a}}{(\frac{V_{0} }{a})^{\frac{2}{3}}}\Big]} &  \mathrm{,0 \leq x \leq a}\\
   \mathrm{ x} & \mathrm{,x>a.}
  \end{cases}
\end{align}
In order to visualise the wormhole geometry, let us consider  $V_{0} = a$. Thus, $V_{0} = a = z= 1.58377$, which gives the values of the constants as $C= 4.16233$ and $D= 0.172416$. The throat radius $(b_{0})$ of the wormhole thus constructed, is given as:
\begin{align}
   \mathrm{ b_{0}\,= r(x=0)= a \Big({\frac{0.44 + \frac{1}{z}0.61}{0.318}} Ai[z]+ {\frac{-0.26 + \frac{1}{z}0.35}{-0.318}} Bi[z] \Big)= a(0.39098)\,= 0.619222\, .}
\end{align}
\noindent Note that the throat radius depends only on $a$ since $z$ is already fixed.
Hence one can choose a $b_0$ which will lead to fixed values of
$a$ and $V_0$.
 
 \noindent The plot below (Figure 2) shows the behavior of $r(x)$ with respect to $x$ for the above values of $V_0$ and $a$. It should be noted that the second derivative of $r(x)$ is discontinuous which is reflected in the fact that the wormhole we constructed has matter only in the region $-a \leq x \leq a$. Beyond $x=a,-a$, the metric is that of flat spacetime. At the boundaries, the curvature tensors  have a discontinuity
 reminiscent of the construction of a `sandwich' spacetime.
\vspace{0.1in}
 \begin{figure}[H]
      \centering
      \includegraphics[scale=0.85]{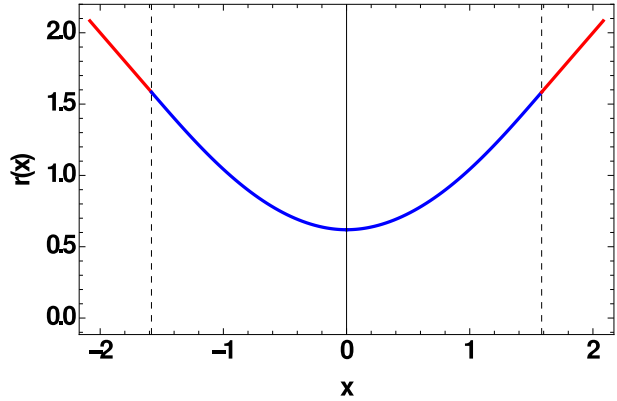}
      \caption{Variation of $r(x)$}
      \label{fig:fig2}
  \end{figure}  
 
\noindent  We can obtain the energy-- momentum tensor corresponding to this wormhole and try to understand the nature of matter contained in the region $-a \leq x \leq a$. From the Einstein equations of General Relativity (using $8 \pi G = c = 1$) and the Einstein tensor for the above-stated line element,
we get (equating $T_{00}= \rho(x)$, $T_{11}= \tau(x)$, $T_{22}= T_{33}= p(x)$, as defined in the frame basis),
\begin{gather}
   \mathrm{ \rho(x) = -\frac{2 r(x) r''(x)+r'(x)^2-1}{r(x)^2} }\\
   \mathrm{ \tau(x)= \frac{r'(x)^2-1}{r(x)^2}}\\
   \mathrm{ p(x) = \frac{r''(x)}{r(x)}.}
\end{gather}
In the above, one needs to use the expression for $r(x)$ obtained in equation (21) for the region $-a \leq x \leq a$. For any possible value of $a$ and $V_0$ the energy density will be a negative quantity. If we take the case $a= V_0 = 1.58377$, then at the throat, $\rho(x=0) = -0.559$ in appropriate units. Therefore, as 
expected  \cite{morris}, exotic matter, violating the Weak Energy Condition (WEC), is required to support this wormhole too.

\section{QUASINORMAL MODES AND BOUND STATES}
\subsection{Quasi-normal modes}
\noindent Quasi-normal modes are a discrete set of complex frequencies which satisfy the boundary conditions of purely outgoing waves at $x \rightarrow \mp \infty$, for an open system \cite{Vishveshwara,cardoso}. To find the QNMs in a given problem
we begin with the
standard $1+1$ dimensional wave equation  in the presence of a potential $V(x)$, i.e. we have
\begin{align}
\mathrm{\frac{\partial^2 \phi}{\partial t^2} -\frac{\partial^2 \phi}{\partial x^2} + V(x) \phi =0 \, .}
\end{align}
Using $\mathrm{\phi (t,x) = e^{-i\omega t} \psi(x)}$,  
the spatial wave function $\psi(x)$ obeys a time-independent Schr\"{o}dinger equation:
\begin{align}
    \mathrm{\frac{d^{2} \psi(x)}{dx^{2}} +(E - V(x)) \psi(x) = 0 \hspace{0.2in} \longrightarrow \hspace{0.2in} \frac{d^{2} \psi(x)}{dx^{2}} + (\omega^2 - V(x)) \psi(x) = 0 \, .}
\end{align}
The coordinate $x$ is a spatial coordinate ranging from $-\infty$ to $+ \infty$ and $E=\omega^{2}$. Time $t$ runs from $0$ to infinity. 
The  equation for $\psi(x)$ needs to be solved with 
proper outgoing boundary conditions at the spatial infinities, 
if $\omega$ is to represent a complex quasi-normal frequency. The corresponding $\phi(x,t)$ is the quasi-normal mode with $\mathrm{\phi(x,t) = e^{-i \omega t} \psi(x)}$. For an asymptotically flat, static spacetime, the potential is positive and satisfies,\\
\begin{align}
      \mathrm{V(x) \rightarrow 0 \hspace{0.3in} as \hspace{0.3in} x \rightarrow \pm \infty}
\end{align}
with the boundary conditions being:\\
\begin{align}
     \mathrm{ \psi \rightarrow e^{-i \omega x}\hspace{0.2in} as\hspace{0.2in} x \rightarrow -\infty \hspace{0.3in}  and \hspace{0.3in} \psi \rightarrow e^{i \omega x}\hspace{0.2in} as\hspace{0.2in} x \rightarrow \infty \, .}
\end{align}
For a black hole spacetime this is physically motivated and makes sense as at the horizon $x \rightarrow -\infty$, there shall only be an ingoing wave, while at spatial infinity $x \rightarrow \infty$ there will only be the outgoing wave
carrying the energy away from the system to infinity. Note that for $\omega$ with a negative imaginary part, the
time part $e^{-i\omega t}$ behaves like a damped oscillation and the spatial part diverges to infinity. Thus,
the meaningful use of the QNMs appears to be through the nature of the temporal part. The spatial part is
evaluated at fixed $x$ (finite) and its equation, when solved, yields the QNMs. 

\subsection{Ferrari-Mashhoon method: a bound state--quasinormal mode correspondence}
\noindent In their 1984 paper, Ferrari and Mashhoon\cite{mashhoon2,ferrari} devised a new analytical technique for finding the QNMs of black hole spacetimes. The method was based on the fact that there is a connection between the QNMs and the bound states of the inverted potentials  \cite{mashhoon1,mashhoon2,ferrari,blome}. Let us begin with the time independent Schr\"{o}dinger equation for finding the bound states ($E <0$) of a potential well defined by $-V(x,p)$, where $p$ is some parameter appearing in the potential.
\begin{align}
    \mathrm{ -\frac{d^{2} \psi}{dx^{2}} - V(x,p) \psi = - E \psi \hspace{0.2in} \longrightarrow \hspace{0.2in} \frac{d^{2} \psi}{dx^{2}} +V(x,p) \psi = E \psi  }
\end{align}
The bound state wavefunction and the corresponding frequency also depends on the parameter $p$, $\psi = \psi(x,p)$ and $\omega = \omega(p)$.\\
Consider the transformation
\begin{align}
   \mathrm{ x \longrightarrow - i x \hspace{0.2in} and \hspace{0.2in} p \longrightarrow  p'}
\end{align}
such that the potential remains invariant
\begin{align}
   \mathrm{ V(x,p) = V(- i x, p').}
\end{align}
The equation $(30)$ under this transformation becomes-
\begin{align}
   \mathrm{ \frac{d^{2} \psi}{dx^{2}} \, \longrightarrow  \,- \frac{d^{2} \psi}{dx^{2}} \hspace{0.1in} ;\, 
   \frac{d^{2} \psi}{dx^{2}} - V(- i x, p') \psi = - E \psi \hspace{0.1in}
     =>  - \frac{d^{2} \psi}{dx^{2}} + V(x , p) \psi = E \psi }
\end{align}
which is nothing but a time-independent Schr\"{o}dinger equation for scattering states from a potential barrier $V( x , p)$. Thus, the method says that if we can find a formula for bound state energy, then the QNMs can be obtained from the bound states just by using the transformation in (31). We will later see that for the triangular potential $p'= -ip$, which is a special case.

\noindent The QNMs found using the correspondence must satisfy the proper boundary conditions. To see this, let us take a look at the Schr\"{o}dinger equation for the potential well when $V = 0$ (i.e. outside the well) given as,
\begin{align}
\mathrm{ \frac{d^{2} \psi}{dx^{2}} - E \psi = 0 \, .}
\end{align}
The solution is,
\begin{align}
 \mathrm{ \psi (x, p) \propto e^{\pm \omega x} \hspace{0.2in};    x \longrightarrow \mp \infty \,.}
\end{align}
If we use the transformation $x \longrightarrow - i x$ then this solution becomes 
\begin{align}
 \mathrm{ \psi \propto e^{\mp i \omega x} \hspace{0.2in} as \hspace{0.2in} x \longrightarrow \mp \infty}
  \end{align}
  which is the boundary condition for QNMs. One can also check this by including the time part and
writing the full wave function $\phi (t,x)$.

\section{BOUND STATES OF THE TRIANGULAR WELL}

\noindent In order to find the bound states ($E < 0$), let us consider, as shown in Figure $1(a)$, three regions: region I for $- \infty < x <-a$, region II for $-a \leq x \leq a$ and region III for  $a < x < \infty$.
\subsection{Solving the Schr\"{o}dinger equation}

\noindent\textbf{Region I:}

\vspace{0.2in}
\noindent The Schr\"{o}dinger equation is given as,
\begin{align}
  \mathrm{ \frac{d^{2} \psi}{dx^{2}} - E \psi = 0 \hspace{0.1in}
   => \hspace{0.1in} \frac{d^{2} \psi}{dx^{2}} - \omega^{2} \psi = 0}
\end{align}
\noindent where we have taken $\frac{2 m}{\hbar^{2}}=1$ and $\omega = \sqrt{E}$. The solution is given by,
\begin{align}
    \mathrm{\psi(x) = A e^{\omega x}}
\end{align}
\noindent with A being a constant. $\psi$ remains finite and tends to zero as $x\,\rightarrow \,-\infty$.\\

\noindent\textbf{Region III:}

\vspace{0.2in}
\noindent The Schr\"{o}dinger equation remains the same as in Region I with the solution being,
\begin{align}
   \mathrm{ \psi(x) = B e^{- \omega x}\, ,}
\end{align}
\noindent B being some arbitrary constant which has to be equal to the constant A for $\psi$ to be symmetric or anti-symmetric. $\psi$ remains finite and tends to zero as $x\,\rightarrow \,\infty$.\\

\noindent\textbf{Region II:}

\vspace{0.2in}
\noindent The equation in this region is given as
\begin{gather}
   \mathrm{\frac{d^{2} \psi}{dx^{2}}} = \begin{cases}\mathrm{\Big(\frac{V_{0} x}{a} + E - V_{0}\Big) \psi} & ;\mathrm{ \,\, 0 \leq x \leq a }\\
   \\
 \mathrm{\Big(-\frac{V_{0} x}{a} + E - V_{0}\Big) \psi} & ; \,\, \mathrm{-a \leq x \leq 0.}
   \end{cases}
\end{gather}
The above equation is the Airy differential equation with its general solution given  as,
\begin{align}
   \mathrm{ \psi(x)}=\begin{cases}
    \mathrm{C\, Ai\Big[\frac{\frac{V_{0} x}{a}+ E-V_{0}}{(\frac{V_{0}}{a})^{\frac{2}{3}}}\Big] + D\, Bi\Big[\frac{\frac{V_{0} x}{a}+ E-V_{0}}{(\frac{V_{0}}{a})^{\frac{2}{3}}}\Big]} & ;\,\,\mathrm{ 0 \leq x \leq a}\\
    \\
     \mathrm{  C Ai\Big[\frac{\frac{-V_{0} x}{a}+ E-V_{0}}{(\frac{V_{0}}{a})^{\frac{2}{3}}}\Big] + D\, Bi\Big[\frac{\frac{-V_{0} x}{a}+ E-V_{0}}{(\frac{V_{0}}{a})^{\frac{2}{3}}}\Big]} & ; \,\,\mathrm{-a \leq x \leq 0}\\
    \end{cases}
\end{align}
\noindent where $C$ and $D$ are constants and $Ai$, $Bi$ are the Airy functions of first and second kind respectively. Following the argument mentioned in Section II for equation (12) here also we take only one root of $(-1)^{1/3}$ i.e. -1 and squaring it gives 1 in the denominator of the argument of the Airy functions for the solution of $-a \leq x \leq 0$.

\subsection{Constructing even states and matching at boundary}
\noindent The even states are constructed from the above solution as follows:
\begin{gather}
    \mathrm{\psi(x) }=\begin{cases}\mathrm{ A e^{\omega x}} & ; \,\,\mathrm{x<-a}\\
     \mathrm{C\, Ai\Big[\frac{\frac{-V_{0} x}{a}+ E-V_{0}}{(\frac{V_{0}}{a})^{\frac{2}{3}}}\Big] + D\, Bi\Big[\frac{\frac{-V_{0} x}{a}+ E-V_{0}}{(\frac{V_{0}}{a})^{\frac{2}{3}}}\Big]} &  ; \,\,\mathrm{ -a \leq x \leq 0}\\
     \\
     \mathrm{C\,Ai\Big[\frac{\frac{V_{0} x}{a}+ E-V_{0}}{(\frac{V_{0}}{a})^{\frac{2}{3}}}\Big] + D\, Bi\Big[\frac{\frac{V_{0} x}{a}+ E-V_{0}}{(\frac{V_{0}}{a})^{\frac{2}{3}}}\Big]} & ; \,\,\mathrm{ 0 \leq x \leq a }\\
     \mathrm{A e^{- \omega x}}& ; \,\, \mathrm{x > a} \, .
     \end{cases}
\end{gather}
\noindent Since the wavefunction and its derivative are continuous over the entire region, we match $\psi$ and $\psi'$ across $x=a$,
\begin{gather}
   \mathrm{ A e^{- \omega a} = \,C\, Ai\Big[\frac{E}{(\frac{V_{0}}{a})^{\frac{2}{3}}}\Big]\, +\,D\, Bi\Big[\frac{E}{(\frac{V_{0}}{a})^{\frac{2}{3}}}\Big]}\\
   \mathrm{ - \omega A e^{-\omega a}= \,\Big(\frac{V_{0}}{a}\Big)^{\frac{1}{3}}\, \Big( C\, Ai'\Big[\frac{E}{(\frac{V_{0}}{a})^{\frac{2}{3}}}\Big]\, +\,D\, Bi'\Big[\frac{E}{(\frac{V_{0}}{a})^{\frac{2}{3}}}\Big]\Big) \, }.
\end{gather}
\noindent Hence, by taking the ratio of equations $(43)$ and $(44)$ we get
\begin{align}
   \mathrm{ \frac{C}{D}} = \mathrm{-\frac{Bi\Big[\frac{E}{(\frac{V_{0}}{a})^{\frac{2}{3}}}\Big]+(\frac{1}{\sqrt{E}})\,(\frac{V_{0}}{a})^{\frac{1}{3}}\,Bi'\Big[\frac{E}{(\frac{V_{0}}{a})^{\frac{2}{3}}}\Big]}{Ai\Big[\frac{E}{(\frac{V_{0}}{a})^{\frac{2}{3}}}\Big]+(\frac{1}{\sqrt{E}})\,(\frac{V_{0}}{a})^{\frac{1}{3}}\,Ai'\Big[\frac{E}{(\frac{V_{0}}{a})^{\frac{2}{3}}}\Big]} \, .}
\end{align}
\noindent Since the coefficients $C$ and $D$ are undetermined, we use another condition. For an even wavefunction, $\psi'(0) = 0$. We have to explicitly apply this condition as our wavefunctions are not inherently symmetric or anti-symmetric. Applying this we get the ratio $\frac{C}{D}$ as
\begin{align}
   \mathrm{ \frac{C}{D} = -\frac{Bi'\Big[\frac{E-V_{0}}{(\frac{V_{0}}{a})^{\frac{2}{3}}}\Big]}{Ai'\Big[\frac{E-V_{0}}{(\frac{V_{0}}{a})^{\frac{2}{3}}}\Big]} \, .}
\end{align}
\noindent Hence the final transcendental equation, the roots of which 
are the even bound state energies, is given as,
\begin{align}
  \mathrm{\frac{Bi\Big[\frac{E}{(\frac{V_{0}}{a})^{\frac{2}{3}}}\Big]+(\frac{1}{\sqrt{E}})\,(\frac{V_{0}}{a})^{\frac{1}{3}}\,Bi'\Big[\frac{E}{(\frac{V_{0}}{a})^{\frac{2}{3}}}\Big]}{Ai\Big[\frac{E}{(\frac{V_{0}}{a})^{\frac{2}{3}}}\Big]+(\frac{1}{\sqrt{E}})\,(\frac{V_{0}}{a})^{\frac{1}{3}}\,Ai'\Big[\frac{E}{(\frac{V_{0}}{a})^{\frac{2}{3}}}\Big]} = \frac{Bi'\Big[\frac{E-V_{0}}{(\frac{V_{0}}{a})^{\frac{2}{3}}}\Big]}{Ai'\Big[\frac{E-V_{0}}{(\frac{V_{0}}{a})^{\frac{2}{3}}}\Big]} \, .}
\end{align}

\subsection{Constructing odd states and matching at boundary}

%\vspace{0.1in}
\noindent The odd states are constructed as follows
\begin{gather}
    \mathrm{ \psi(x)} =\begin{cases}\mathrm{- A e^{\omega x} }& \mathrm{,x<-a}\\
   \mathrm{ - C\, Ai\Big[\frac{\frac{-V_{0} x}{a}+ E-V_{0}}{(\frac{V_{0}}{a})^{\frac{2}{3}}}\Big] - D\, Bi\Big[\frac{\frac{-V_{0} x}{a}+ E-V_{0}}{(\frac{V_{0}}{a})^{\frac{2}{3}}}\Big]} &   \mathrm{,-a \leq x \leq 0}\\
    \\
    \mathrm{ C\,Ai\Big[\frac{\frac{V_{0} x}{a}+ E-V_{0}}{(\frac{V_{0}}{a})^{\frac{2}{3}}}\Big] + D\, Bi\Big[\frac{\frac{V_{0} x}{a}+ E-V_{0}}{(\frac{V_{0}}{a})^{\frac{2}{3}}}\Big]} & \mathrm{,0 \leq x \leq a }\\
    \mathrm{A e^{- \omega x}}& \mathrm{,x > a \, .}
     \end{cases}
\end{gather}
\noindent Proceeding in a way similar to the case for even states, we match the wavefunction and its derivative at the boundary to get,
\begin{align}
   \mathrm{ \frac{C}{D}= -\frac{Bi\Big[\frac{E}{(\frac{V_{0}}{a})^{\frac{2}{3}}}\Big]+(\frac{1}{\sqrt{E}})\,(\frac{V_{0}}{a})^{\frac{1}{3}}\,Bi'\Big[\frac{E}{(\frac{V_{0}}{a})^{\frac{2}{3}}}\Big]}{Ai\Big[\frac{E}{(\frac{V_{0}}{a})^{\frac{2}{3}}}\Big]+(\frac{1}{\sqrt{E}})\,(\frac{V_{0}}{a})^{\frac{1}{3}}\,Ai'\Big[\frac{E}{(\frac{V_{0}}{a})^{\frac{2}{3}}}\Big]} \, .}
\end{align}
\noindent Here, since we have odd states, to determine the ratio $\frac{C}{D}$ we evaluate $\psi(0) = 0$ which gives,
\begin{align}
     \mathrm{ \frac{C}{D} = -\frac{Bi\Big[\frac{E-V_{0}}{(\frac{V_{0}}{a})^{\frac{2}{3}}}\Big]}{Ai\Big[\frac{E-V_{0}}{(\frac{V_{0}}{a})^{\frac{2}{3}}}\Big]} \, .}
\end{align}
\noindent Hence the final transcendental equation for odd bound state energies becomes,
\begin{align}
   \mathrm{ \frac{Bi\Big[\frac{E}{(\frac{V_{0}}{a})^{\frac{2}{3}}}\Big]+(\frac{1}{\sqrt{E}})\,(\frac{V_{0}}{a})^{\frac{1}{3}}\,Bi'\Big[\frac{E}{(\frac{V_{0}}{a})^{\frac{2}{3}}}\Big]}{Ai\Big[\frac{E}{(\frac{V_{0}}{a})^{\frac{2}{3}}}\Big]+(\frac{1}{\sqrt{E}})\,(\frac{V_{0}}{a})^{\frac{1}{3}}\,Ai'\Big[\frac{E}{(\frac{V_{0}}{a})^{\frac{2}{3}}}\Big]} = \frac{Bi\Big[\frac{E-V_{0}}{(\frac{V_{0}}{a})^{\frac{2}{3}}}\Big]}{Ai\Big[\frac{E-V_{0}}{(\frac{V_{0}}{a})^{\frac{2}{3}}}\Big]} \, .}
\end{align}

\subsection{Finding even and odd eigenvalues and eigenfunctions}
\noindent In order to obtain representative values for the energies 
and show their respective wave functions, we consider the following special cases: $\frac{V_{0}}{a}>1$, $\frac{V_{0}}{a} <1$ and $\frac{V_{0}}{a} = 1$. The energy eigenvalues are listed below in Table I with the normalised wavefunctions shown in Figure 3 for $V_{0} =5\,,a=2$. \\

\begin{figure}[H]
 \centering
 \begin{subfigure}[t]{0.47\textwidth}
 	\centering
 	\includegraphics[width=\textwidth]{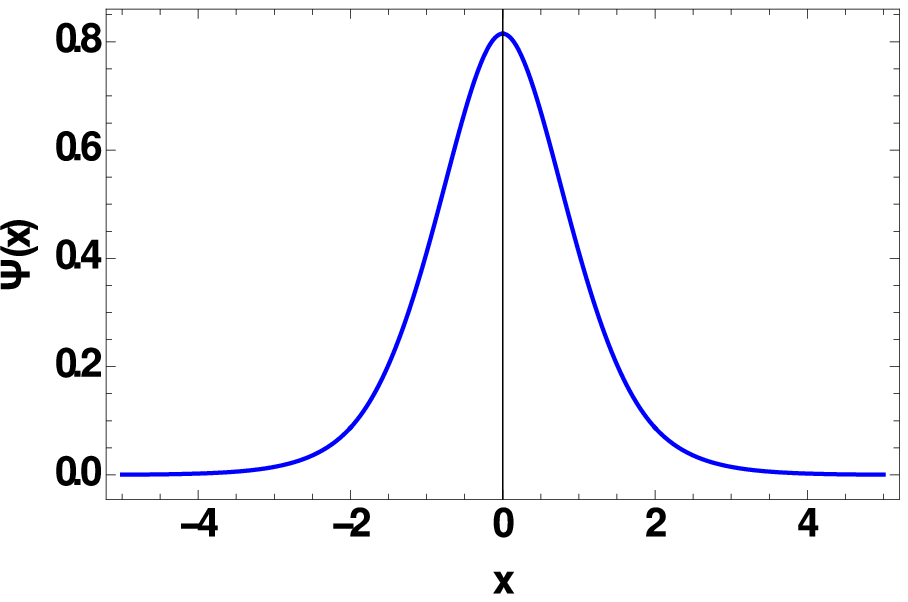}
 	\caption{ Even state of $V_{0} = 5, a = 2$}
 	\label{fig:fig(a)}
 \end{subfigure}
 \hspace{0.15in}
 \begin{subfigure}[t]{0.47\textwidth}
 	\centering
 	\includegraphics[width=\textwidth]{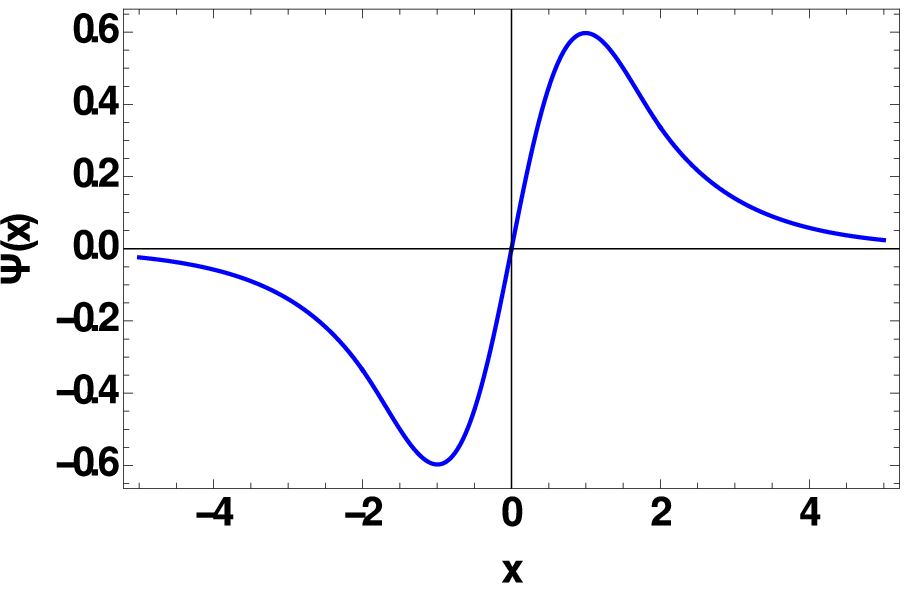}
 	\caption{Odd state of $V_{0} = 5, a = 2$}
 	\label{fig:fig(b)}
 \end{subfigure}
\caption{ Normalised wavefunction for $V_{0} = 5, a=2$. }
 \label{fig:fig3}
\end{figure}

\begin{table}[H]
\begin{center}
\begin{tabular}{||c|c|c||}
 \toprule[0.8pt]
    & Even state & Odd state\\ [1ex]
    \hline \hline
   $V_{0} = a =2$ & 0.99364 & 0.000266614\\ [1ex]
   
    $V_{0} = 5, a=2$ & 3.12575 & 0.775377\\ [1ex]
  
    $V_{0} = 2, a=5$ & 0.254455 & 0.731984\\ [1ex]
   
    $V_{0} = a = 1$  & 0.173819 & No odd state\\ [1ex]
    \bottomrule[0.8pt]
 \end{tabular}
 \caption{\label{tab:table-1} Energy eigenvalues for even and odd states.}
\end{center}
\end{table}

%\vspace{0.15in}
\section{FINDING THE QUASI-NORMAL MODES}
\noindent 
Let us now move on towards finding the QNMs in the barrier potential. Using the Ferrari-Mashhoon idea we define a transformation which will keep the potential unchanged.
This transformation is given as,
\begin{gather}
    \mathrm{x\,\longrightarrow \, - i x \, .}
\end{gather}
which changes the sign of the second derivative term in the time-indepenedent 
Schrodinger equation. This results in the well becoming a barrier.
\noindent In our case, the potential remains invariant since 
$a \longrightarrow -i a$.  The solutions for the different regions can be 
obtained from the bound states by simply imposing the transformation mentioned above. One may find 
the QNMs directly by solving the Schr\"{o}dinger equation for a barrier with proper QNM boundary conditions. However,  since our potential is invariant under the transformation $(52)$ one can also get the QNMs by making use of the Ferrari-Mashhoon method. 
It is important to note that while doing so, one must keep in mind that we need to consider only $(i)^{\frac{1}{3}}= - i$ as the other roots of $(i)^{1/3}$ can be absorbed by considering $e^{\frac{\pm 2 i \pi}{3}}$ in the argument of the Airy function and remembering that they do not give any new solution \cite{DLMF}. Thus the even set of solutions in the three regions of the barrier (Fig. 1b) become
\begin{gather}
   \mathrm{\psi(x) }=\begin{cases}\mathrm{ A e^{-i\omega x}} & ; \,\,\mathrm{ x<-a}\\
     \mathrm{C\, Ai\Big[\frac{\frac{V_{0} x}{a}- E+V_{0}}{(\frac{V_{0}}{a})^{\frac{2}{3}}}\Big] + D\, Bi\Big[\frac{\frac{V_{0} x}{a}- E+V_{0}}{(\frac{V_{0}}{a})^{\frac{2}{3}}}\Big]} &   ; \,\, \mathrm{-a \leq x \leq 0}\\
     \\
    \mathrm{ C\,Ai\Big[\frac{\frac{-V_{0} x}{a}- E+V_{0}}{(\frac{V_{0}}{a})^{\frac{2}{3}}}\Big] + D\, Bi\Big[\frac{\frac{-V_{0} x}{a}- E+V_{0}}{(\frac{V_{0}}{a})^{\frac{2}{3}}}\Big]} & ; \,\,\mathrm{0 \leq x \leq a} \\
     \mathrm{A e^{i\omega x} }& ; \,\, \mathrm{x > a} \, .
     \end{cases}
\end{gather}
\noindent For the odd set of wavefunctions there will be a negative sign in the amplitude of $\psi(x)$ in the region $-\infty <x <0$. The quasi-normal modes for the even and odd states are obtained by solving the transcendental equations $(55)$ and $(56)$ (see below) respectively which have been obtained by imposing the transformation on equations $(47)$ and $(51)$. In the argument of the Airy functions of equations (47) and (51) we impose $a \rightarrow -ia$ which gives $(i)^{2/3}$. As mentioned earlier we take $(i)^{1/3} = -i$ and squaring it gives $-1$. Thus, we have,
\begin{align}
    \mathrm{\frac{E}{(\frac{V_{0}}{a})^{\frac{2}{3}}} \xrightarrow{a \rightarrow -ia}     \frac{-E}{(\frac{V_{0}}{a})^{\frac{2}{3}}}.}
\end{align}
Also the factor of $(\frac{V_{0}}{a})^{1/3}$ provides the additional factor of $i$ multiplied with $\sqrt{E}$. Hence we get the following transcendental equations as
\begin{align}
    \mathrm{\frac{Bi\Big[\frac{-E}{(\frac{V_{0}}{a})^{\frac{2}{3}}}\Big]+(\frac{1}{i \sqrt{E}})\,(\frac{V_{0}}{a})^{\frac{1}{3}}\,Bi'\Big[\frac{-E}{(\frac{V_{0}}{a})^{\frac{2}{3}}}\Big]}{Ai\Big[\frac{-E}{(\frac{V_{0}}{a})^{\frac{2}{3}}}\Big]+(\frac{1}{i \sqrt{E}})\,(\frac{V_{0}}{a})^{\frac{1}{3}}\,Ai'\Big[\frac{-E}{(\frac{V_{0}}{a})^{\frac{2}{3}}}\Big]} = \frac{Bi'\Big[\frac{V_{0}-E}{(\frac{V_{0}}{a})^{\frac{2}{3}}}\Big]}{Ai'\Big[\frac{V_{0}-E}{(\frac{V_{0}}{a})^{\frac{2}{3}}}\Big]} }
\end{align}
\begin{align}
      \mathrm{ \frac{Bi\Big[\frac{-E}{(\frac{V_{0}}{a})^{\frac{2}{3}}}\Big]+(\frac{1}{i \sqrt{E}})\,(\frac{V_{0}}{a})^{\frac{1}{3}}\,Bi'\Big[\frac{-E}{(\frac{V_{0}}{a})^{\frac{2}{3}}}\Big]}{Ai\Big[\frac{-E}{(\frac{V_{0}}{a})^{\frac{2}{3}}}\Big]+(\frac{1}{i \sqrt{E}})\,(\frac{V_{0}}{a})^{\frac{1}{3}}\,Ai'\Big[\frac{-E}{(\frac{V_{0}}{a})^{\frac{2}{3}}}\Big]} = \frac{Bi\Big[\frac{V_{0}-E}{(\frac{V_{0}}{a})^{\frac{2}{3}}}\Big]}{Ai\Big[\frac{V_{0}-E}{(\frac{V_{0}}{a})^{\frac{2}{3}}}\Big]} }
\end{align}
for even and odd QNMs, respectively.

\subsection{Solving the transcendental equations}
\noindent In order to find the roots of the transcendental equations (55) and (56) we do a contour plot in {\em Mathematica 10.0} \cite{mathematica} by equating the real and imaginary parts of LHS and RHS so that the intersection points will give us the initial guess for the quasinormal frequencies. Next we use the {\em FindRoot} command in {\em Mathematica 10.0} 
to explicitly find the numerical value of the root using an initial guess obtained from the contour plot. In the table below (Table II) some of the QNM frequencies are mentioned for different values of the parameters in the potential. The plot in Figure 4 shows the even and odd QNM wavefunctions for the $V_{0} =a = 1.58377$ which we had obtained in Section II. Thus the QNMs shown in Figure 4 will also be the QNMs for a wormhole geometry under scalar perturbation for the $m=0$ mode.
 
\begin{table}[H]
\begin{center}
\begin{tabular}{||c|c|c||}
     \toprule[0.8pt]
     & Even states & Odd states \\
     \hline \hline
    $V_{0} = a = 2$ & $1.37818 - i\, 0.377625$ & $ 1.84536 - i\, 0.916893$ \\ [1ex]
   % & & $2.58148 - i\, 1.21217$\\
   \midrule[0.4pt]
    $V_{0} = 5, a=2$ & $2.08866 - i\, 0.323045$ & $2.32235 - i\, 0.723801$\\ [1ex]
    & $2.92932 - i\, 0.947045$ & $3.62696 -  i\, 1.17254$\\ [1ex]
  \midrule[0.4pt]
    $V_{0} =2, a=5$ & $1.2607 - i\, 0.184607$ & $1.2408 - i\, 0.263773 $\\ [1ex]
    & $1.43577 - i\, 0.252766 $ & $1.66907 - i\, 0.367693 $\\ [1ex]
    \midrule[0.4pt]
    $V_{0} = a = 1.58377$ & $1.21957 - i\, 0.58865$ & $2.03544 - i\, 1.40477$\\ [1ex]
    \bottomrule[0.8pt]
\end{tabular}
\caption{\label{tab:table-2}Values of  $\omega_{QNM}$ for different potentials.}
\end{center}
\end{table}
%\noindent The wavefunctions for $V_{0}$ and $a$  with each root of $i$ is shown in figure.

\begin{figure}[H]
 \centering
 \begin{subfigure}[t]{0.47\textwidth}
 	\centering
 	\includegraphics[width=\textwidth]{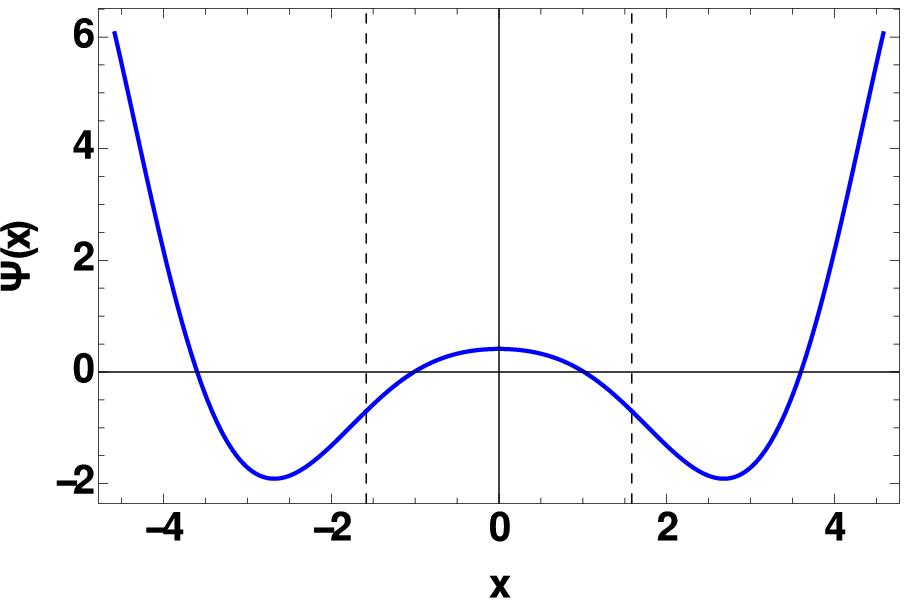}
 	\caption{  Even QNM}
 	\label{fig:fig(a)}
 \end{subfigure}
 \hspace{0.2in}
 \begin{subfigure}[t]{0.49\textwidth}
 	\centering
 	\includegraphics[width=\textwidth]{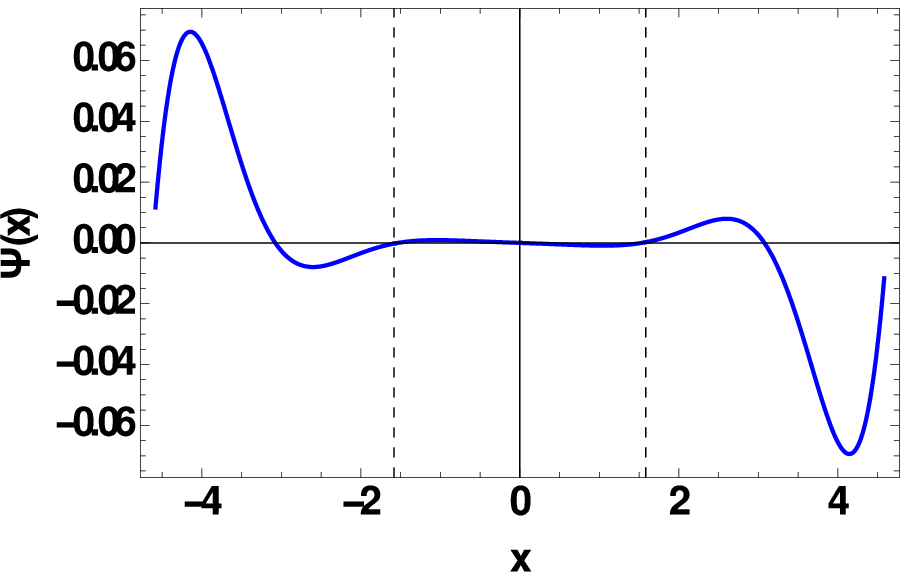}
 	\caption{  Odd QNM}
 	\label{fig:fig(b)}
 \end{subfigure}
 \caption{ QNMs for $V_{0} = a = 1.58377$.}
 \label{fig:fig4}
\end{figure}

\subsection {The time domain profile}

\noindent The time domain profile of a wave function satisfying a given wave equation, shows its evolution with time at a particular value of spatial position. The time domain profile can also be used to obtain the QNM frequencies as described in \cite{price} and \cite{konoplya}. It is found by directly integrating the wave-like equation (26) and observing the behavior of the field, at a particular spatial coordinate, with time. The entire time domain profile comprises of the initial transient phase, followed by the QNM ringing (during the ringdown phase) and, finally, the late time behavior.  In order to generate such a profile we first write equation (26) in light cone coordinates $du = dt - dx $ and $dv = dt + dx$. This gives,
\begin{align}
  \mathrm{ \Big( 4 \frac{\partial^{2}}{\partial u \partial v} + V(u,v) \Big) \phi = 0.}
\end{align}
The time evolution operator in light cone coordinates can be written as,
\begin{align}
    \begin{split}
      \mathrm{ exp\Big(h \frac{\partial}{\partial t}\Big) = exp\Big( h\frac{\partial}{\partial u} + h \frac{\partial}{\partial v} \Big) 
    = exp \Big(h \frac{\partial}{\partial u}\Big) + exp\Big( h\frac{\partial}{\partial v}\Big) -1}\\ \mathrm{+ \frac{h^{2}}{2} \Big( exp\Big( h\frac{\partial}{\partial u}\Big) + exp\Big(h \frac{\partial}{\partial v} \Big) \Big) \frac{\partial^{2}}{\partial u \partial v} + ....}
    \end{split}
\end{align}

\noindent When this operates on $\phi$ and incorporating (57), we get the discretised form of the differential equation with step size $h$ given as: 
\begin{align}
    \mathrm{\phi(u+h,v+h) = \phi(u+h,v) + \phi(u,v+h) - \phi(u,v) - \frac{h^{2}}{8} V(u,v) (\phi(u,v+h)+ \phi(u+h,v))+.....}
\end{align}
We integrate along each cell of a $uv$ grid bounded by two null surfaces $u=u_{0}$ and $v=v_{0}$ with $h=0.1$. The initial conditions on the $v=0$ line is a Gaussian profile centered at $u=10$, i.e. $\psi(u,0)= e^{\frac{-(u-10)^{2}}{100}}$. On the $u=0$ line $\psi$ is taken as a constant, $\psi(0,v)= constant$, the value of which is determined by the value of $\psi (0,0)$. The field is obtained in the region $0<u<150$ and $0<v<150$. We then obtain $\phi(t)$ for a particular value of the 
spatial coordinate. This is plotted in a logarithmic scale as a function of time. The scheme is elegantly described in \cite{konoplya}.  \\

\vspace{0.1in}
\begin{figure}[H]
 \centering
\begin{subfigure}[t]{0.465\textwidth}
  \centering
	\includegraphics[width=\textwidth]{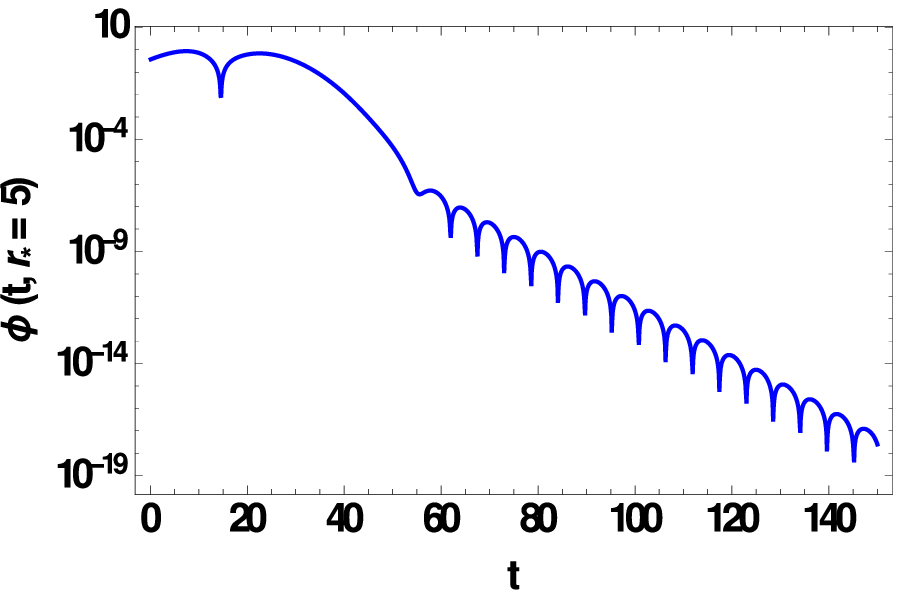}
 	\caption{$V_0 = 0.3412$ and $a = 3.412$}
 	\label{fig:fig(a)}
 \end{subfigure}
 \hspace{0.3in}
 \begin{subfigure}[t]{0.47\textwidth}
 	\centering
 	\includegraphics[width=\textwidth]{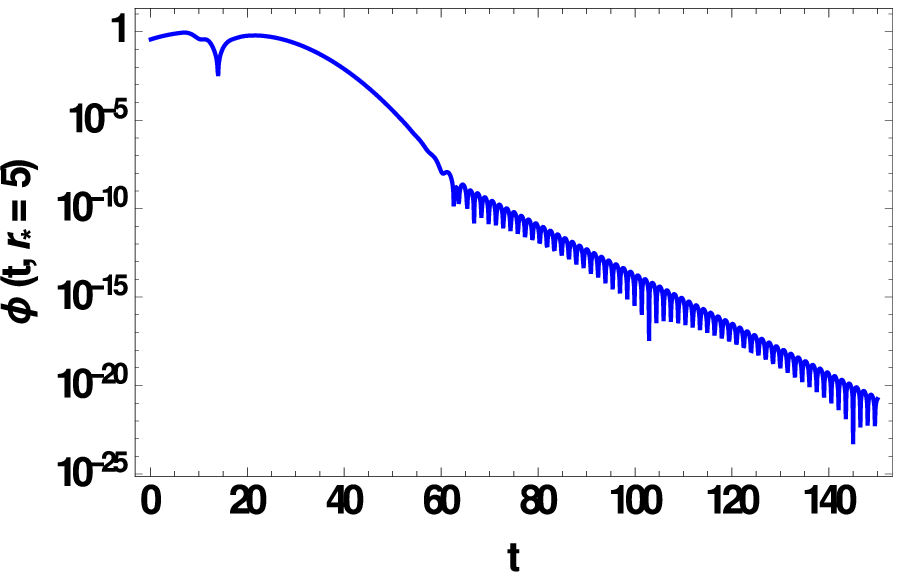}
 	\caption{$V_0 = 5$ and $a = 2$}
 	\label{fig:fig(b)}
 	\end{subfigure}
 	\caption{Time domain profiles showing quasi-normal ringing.}
 \label{fig:fig7}
 \end{figure}

\begin{table}[H]
\begin{center}
\begin{tabular}{|| b{1.5cm}| b{1.5cm} | b{3.5cm} | b{3.55cm}||}
     \toprule[1pt]
     $V_0$ & a & $\omega$ (analytical) & $\omega$ (Prony)\\ [1ex]
    \hline \hline
     $0.3412$ & $3.412$ & 0.56607 -i 0.27322 & 0.566933 -i 0.272709\\  [1ex]
    %\hline
     $1.5837$ & $1.5837$ & 1.21957 -i 0.58865 & 1.22222 -i 0.587214\\ [1ex]
    %\hline
    $7.3512$ & $0.73512$ & 2.62748 -i 1.2682 & 2.62767 -i 1.26129\\ [1ex]
    \midrule[0.5pt]
    $1$ & $10$ & 0.95141 -i 0.081617 &  0.951777 -i 0.0831705\\ [1ex]
    %\hline
    $2$ & $5$ & 1.2607 -i 0.184607 & 1.266 -i 0.187367\\ [1ex]
   % \hline
    $2$ & $2$ & 1.37818 -i 0.377625 & 1.37844 -i 0.377593\\ [1ex]
    %\hline
    $4$ & $4$ & 1.9608 -i 0.247561 & 1.96429 -i 0.249405\\ [1ex]
   % \hline
    $5$ & $2$ & 2.08866 -i 0.323045 & 2.08925 -i 0.325319\\ [1ex]
    \bottomrule[0.8pt]
\end{tabular}
\caption{\label{tab:table-2}Values of QNM frequencies for different $V_0$ and $a$.}
\end{center}
\end{table}
\noindent Figure 5 shows the time-domain profiles
obtained by integrating the wave equation with a triangular barrier potential
using the method mentioned above. We observe that the QNM ringing is quite prominent 
and the amplitude decays with time during this phase. The values of the dominant frequency extracted from the time domain profile matches quite well with the fundamental QNM frequency obtained analytically for different $V_0$ and $a$, as is indicated in Table III. The extraction of frequencies from the time domain profile is done by applying the Prony fitting technique where the QNM spectrum is written as a superposition of damped exponentials \cite{konoplya}. Even though our potential has a derivative discontinuity, we still found the fundamental QNM frequency dominating the time evolution of the field \cite{nollert}.
\noindent For the first three values in Table III, we have taken values of $V_0$ and $a$ such that relation (18) is satisfied, giving us the time domain profile for scalar wave propagation
in the `sandwich' wormhole constructed in section II and the corresponding QNM frequencies. The plot of Fig. 5(a) clearly shows the decaying nature of the scalar field during the QNM ringing phase. Fig 5(b) shows the time domain profile for other values of the $V_0$ and $a$
(not obeying (18)).  The plot in Figure 5(a) demonstrates that the wormhole we have constructed is stable 
w.r.t. scalar perturbations. We can also generalize our comment and state that any spacetime having a perturbation potential in the form of a triangular barrier is likely to be 
be stable under scalar perturbations.

\subsection{Variation of the QNMs with the wormhole throat radius}
\noindent The QNM frequencies have the special characteristic that they depend only on the parameters of the source, like for a black hole the only parameters are its mass, charge and spin. In our wormhole scenario, the parameters are the potential height $V_{0}$ and the width of the potential barrier $2a$ which in turn control the throat radius $b_{0}$ of the wormhole constructed by us as given in equation (22). Thus, in order to observe how the QNM frequencies change for a real physical wormhole (assuming wormholes exist in nature) we need to observe the dependence of the frequency and damping time of the signal on the throat radius. For simplicity, let us consider only the fundamental modes of the even states. If we take the throat radius in units of solar mass $(M_{\odot})$, we get the $\omega_{QNM}$ in units of $(M_{\odot}^{-1})$. In order to convert this into Hz, we need to multiply a factor of $c^{3}/G M_{\odot}$ where $G$ is the gravitational constant, $c$ being the speed of light in vacuum and $M_{\odot}$ being the mass of Sun. The following table (Table IV) lists some of the values of the frequency $\nu = \omega_{re}/2 \pi$ and damping time $\tau / 2 \pi = 1/\mid \omega_{im} \mid$ with the corresponding $b_{0}$.
\begin{table}[H]
\begin{center}
\begin{tabular}{||c | c |c | c | c||}
     \toprule[0.8pt]
     $ b_{0}$  $(M_{\odot})$ & a$ (M_{\odot})$ & $V_0$ $(M_{\odot}^{-2})$ & $\nu $(KHz) & $\tau/(2\pi)$ ($1\times10^{-5}$ sec)\\ [1.5ex]
     \hline \hline
     $0.5$ & 1.2788 & 2.4290 & $48.92$ & $0.674$\\ [1ex]
     $0.8$ & 2.0461 & 0.9488 & $30.57$ & $1.078$ \\ [1ex]
    
    $1$ & 2.5576 & 0.6072 & $24.46$ & $1.348$\\[1ex]
    %\hline
    $1.2$ & 3.0692 & 0.4217& $20.38$ & $1.617$\\[1ex]
   % \hline
    $1.5$ & 3.8365 & 0.2698 & $16.31$ & $2.022$\\[1ex]
   % \hline
    $1.8$ & 4.6038 & 0.1874 & $13.59$ & $2.426$\\[1ex]
   % \hline
    $2$ & 5.1153 & 0.1518 & $12.23$ & $2.696$\\[1ex]
   % \hline
    $2.5$ & 6.3941 & 0.0971 &  $9.78$ & $3.37$\\[1ex]
    \bottomrule[0.8pt]
\end{tabular}
\caption{\label{tab:table-2} Variation of frequency and damping time with throat radius.}
\end{center}
\end{table}

\begin{figure}[H]
 \centering
\begin{subfigure}[t]{0.42\textwidth}
  \centering
	\includegraphics[width=0.85\textwidth]{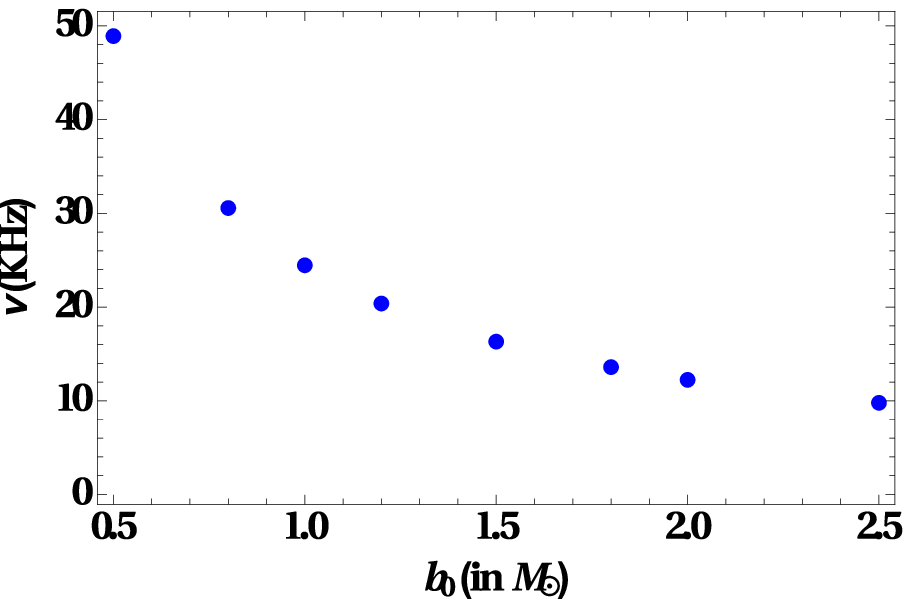}
 	\caption{Variation of frequency with throat radius.}
 	\label{fig:fig(a)}
 \end{subfigure}
 \hspace{0.05in}
 \begin{subfigure}[t]{0.42\textwidth}
 	\centering
 	\includegraphics[width=0.85\textwidth]{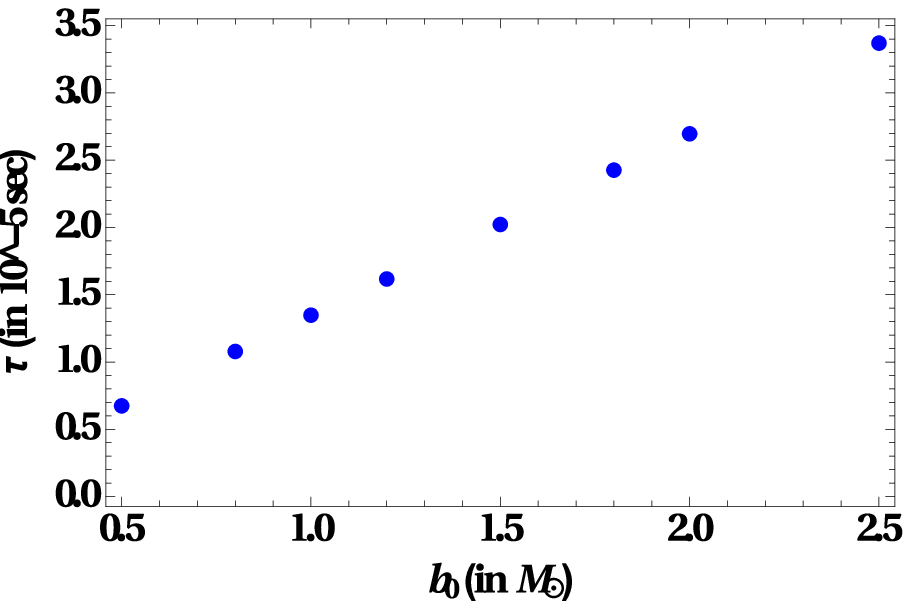}
 	\caption{Variation of damping time (scaled by a factor of $1/(2\pi)$) with throat radius.}
 	\label{fig:fig(b)}
 	\end{subfigure}
 	\caption{QNMs corresponding to even state}
 \label{fig:fig7}
 \end{figure}

\noindent We observe that as the throat radius increases, the frequency of the wave decreases and the damping time increases (Figure 6 (a, b)). This can be interpreted in the following way: if we assume that the wormhole is produced as a result of the  merger of two compact bodies (merger of exotic compact objects has been studied in \cite{cardoso1}, \cite{krishnendu}) then, as the throat radius of the remnant wormhole increases, the frequency of the emitted wave decreases. As and when LIGO sensitivity is enhanced it may be possible to detect QNM signals, in future detections, from macroscopic wormholes with large throat radius \cite{Ligo}, \cite{enhanced ligo}. 

\subsection{ Approximating known potentials with triangular barriers}

\noindent Most of the perturbation potentials that arise 
in astrophysical scenarios are complicated and cannot be solved analytically. 
They can be solved only using numerical techniques. In order to get 
better analytic insight 
on characteristic features such as QNMs, the perturbation potentials 
may be approximated by known exactly solvable potentials like the rectangular barrier \cite{rectangular barrier}, step barrier or the Poschl- Teller \cite{ferrari}. In a similar way, since the triangular barrier is also an exactly solvable model, it may be used to approximate other potentials with which it
has a resemblance in profile (barrier features). Perturbation potentials 
for wormholes are usually of the single or double barrier types. A prominent example is the Ellis--Bronnikov spacetime \cite{ellis}. In the 
approach just mentioned, we can find the scalar QNMs of the Ellis--Bronnikov wormhole \cite{kim,konoplya1,kunz} by approximating it with a triangular barrier.\\
The line element and the effective potential for massless, 
scalar wave propagation in the Ellis--Bronnikov geometry are given by,
\begin{gather}
  \mathrm{ ds^2 = -dt^2 + \frac{dr^2}{1-\frac{b_0^2}{r^2}} + r^2 ( d\theta^2 + sin^2\theta d\phi^2) }\\
   \mathrm{ V_{eff} = \frac{\ell(\ell+1)}{(r_*^2+b_0^2)} + \frac{b_0^2}{(r_*^2+b_0^2)^2}}
\end{gather}
with $\ell$ being related to angular momentum, $r^2 = r_*^2+b_0^2$, where `r' is the radial coordinate, `$r_*$' is the tortoise coordinate and $b_0$ is the throat radius of the wormhole. 
In order to approximate the Ellis--Bronnikov perturbation potential with the triangular barrier we need to define the parameters $a$ and $V_0$ 
(for the triangular barrier) appropriately. The $V_0$ is set equal to the peak of the Ellis--Bronnikov perturbation potential at $r_*=0$ and the value of $a$ is set equal to the full width at half maximum of the curve representing the Ellis--Bronnikov potential (see Figure 7). A similar matching technique for a rectangular potential can be found in the appendix of \cite{kerr WH}.
The spatial coordinate $`x$' in the case of triangular barrier acts as the tortoise coordinate and is similar to $r_*$. The following table (Table V) shows the fundamental QNM (least damped) frequencies obtained both by a numerical technique and by the approximation with the triangular barrier. It also mentions the values of $V_0$ and $a$ used in the calculations. The value of throat radius is taken as unity.

 \begin{figure}[H]
      \centering
      \includegraphics[scale=0.85]{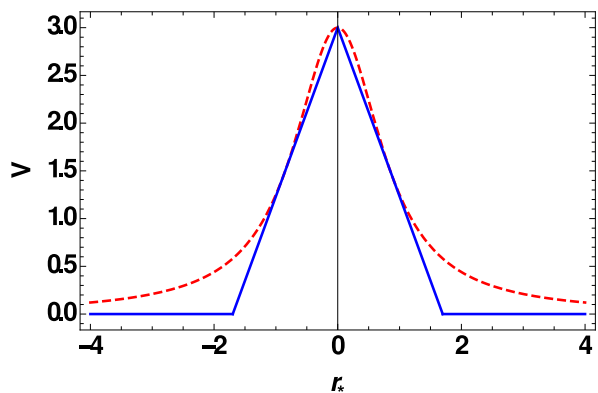}
      \caption{Fitting the Ellis-Bronnikov scalar perturbation potential (dashed line) with the triangular barrier (blue line) for $\ell = 1$ mode.}
      \label{fig:fig2}
  \end{figure} 
  
\noindent To find the numerical results we have used the direct integration approach of Chandrasekhar and Detweiler \cite{Chandrasekhar}. The differential equation (27) is directly integrated with the proper boundary conditions to obtain the QNMs \cite{paolo}. For Ellis--Bronnikov  geometry, the potential is symmetric about $r_*=0$ so the solutions can be divided into symmetric and anti-symmetric types. A similar treatment can be found in \cite{aneesh}. Hence the initial conditions for the two classes of QNMs will be: $\psi(0)=0$ for antisymmetric solutions and $\psi'(0)=0$ for symmetric solutions. We will only consider the symmetric case which has smaller damping. We have followed the procedure for direct integration as given in \cite{aneesh}. The wave function is suitably expanded upto finite but arbitrary orders both near the 
throat and at spatial infinity. It is then substituted in the differential equation and 
evolved from both the throat and spatial infinity towards an in-between arbitrary point. 
Matching the wave function and its derivatives gives the QNMs.\\

\begin{table}[H]
\begin{center}
\begin{tabular}{||c|c|c|c|c||}
     \toprule[0.8pt]
     $\ell$ & $V_0$ & $a$ & $\omega$ (numerical) & $\omega$ (approximation)\\ [1ex]
     \hline \hline
     1 &  3 & 1.69795 & 1.57234 -i 0.529818 & 1.68343 -i 0.435517\\ [1ex]
    % \hline
     2 & 7 & 1.86257 & 2.54627 -i 0.512869 & 2.44744 -i 0.350497\\ [1ex]
    % \hline
     3 & 13 & 1.92461 & 3.53387 -i 0.507184 & 3.22302 -i 0.462328\\ [1ex]
    % \hline
     4 & 21 & 1.95296 & 4.5266 -i 0.504649 & 4.42191 -i 0.436661\\ [1ex]
     %\hline
     5 & 31 & 1.96801 &  5.52182 -i 0.503319 & 5.23639 -i 0.43561\\ [1ex]
    % \hline
     6 & 43 & 1.97688 & 6.51846 -i 0.502545 & 6.45926 -i 0.554812\\ [1ex]
     %\hline
     7 & 57 & 1.98253 & 7.51595 -i 0.502061 & 7.26488 -i 0.463475\\ [1ex]
    % \hline
     8 & 73 & 1.98635 & 8.51401 -i 0.501743 & 8.05989 -i 0.537328\\ [1ex]
     %\hline
     9 & 91 & 1.98904 & 9.51247 -i 0.501526 & 9.309 -i 0.517387\\ [1ex]
    % \hline
     10 & 111 & 1.99101 &  10.5112 -i 0.501373 & 10.1153 -i 0.517864\\ [1ex]
    \bottomrule[0.8pt]
\end{tabular}
\caption{\label{tab:table-2}Values of QNM frequencies for the Ellis-Bronnikov wormhole obtained by approximating with a triangular barrier.}
\end{center}
\end{table}

\noindent It can be seen from Table V that the values obtained using the triangular barrier approximation are quite close to the values obtained by the
above-stated numerical method.
Thus, the triangular barrier can indeed be used to approximate 
similar smooth potentials which may arise as perturbation potentials in different 
spacetimes.

\section{SUMMARY AND DISCUSSION}

\noindent Let us now briefly summarise our results.
\noindent A primary goal of our work as reported in this article was to find the scalar quasi-normal modes of a finite, symmetric triangular barrier potential. 
This problem, though exotic, has not been dealt with before in the literature.
The solutions involve the Airy functions and the QNMs are hidden in
a transcendental equation involving the Airy functions and their derivatives.
A correspondence known due to early work by Ferrari and Mashhoon 
may be used to find
the transcendental equation for the QNMs from those in the bound state problem 
(triangular well). 
We first find the condition for bound states (and the bound state energies and wave-functions) of the triangular well
and use it to find the QNMs using a transformation. It is useful to note that one can also find the
QNMs completely independently, without referring to the bound state problem at all. The complete time domain profiles are then obtained by numerically integrating the
scalar wave equation. We extract the QNMs from the time domain profiles and compare
with the analytically obtained values. The agreement is quite good.

\noindent Further, to provide a context for the triangular barrier, we have constructed a 
`sandwich' wormhole geometry whose effective potential under scalar wave propagation will be a finite, symmetric triangular barrier, for the $m=0$ mode. The wormhole, thus constructed, has matter, with negative energy density, sandwiched over a finite region between two flat spacetimes. We find the QNMs for this
geometry and also show, through the time-domain profiles, that
this wormhole has a decaying late-time tail following the quasi-normal ringdown and is thus
stable. 

\noindent Since wormholes necessarily have perturbation potentials which are
single or double barriers, we try to approximate one such case using the
triangular barrier. We take the scalar perturbations of the Ellis-Bronnikov wormhole and model the barrier potential with a
properly parametrised triangular barrier. A comparison of the QNMs 
for the smooth potential with those for the triangular barrier seems
to be reasonably good. 

\noindent In most realistic cases in gravitational wave or black hole physics the
effective potentials are far more complicated. In order to find QNMs for such scenarios, numerical approaches are extensively employed. There are various numerical techniques available for finding QNMs. These include the direct integration method \cite{Chandrasekhar},\cite{paolo} due to Chandrasekhar-Detweiler, 
the Prony fit method ( a frequency extraction technique 
based on the fact that the time domain profile is dominated, in most cases, by the fundamental QNM 
) \cite{konoplya},\cite{prony1}, \cite{prony2}, \cite{prony3}, the continued fraction method by Leaver \cite{CF}, to list a few. Generically, many of these potentials are 
barrier type (particularly those for wormholes) and may be
modeled using rectangular \cite{rectangular barrier} or triangular barriers, in appropriate limits. Further, double triangular barriers (similar to the double rectangular ones)
may be used  to model the interesting phenomenon of echoes which has generated 
quite some interest and activity in recent times \cite{kerr WH}.

\section{\bf CONCLUSION}
\noindent  In conclusion, we have succeeded in adding a new exactly solvable problem to the list of analytic results (for QNMs)
mentioned in \cite{visser}. We have also shown that the triangular barrier can arise (i) as a scalar perturbation
potential for a specially constructed wormhole geometry or (ii) as an 
approximation to the scalar perturbation potential in the 
Ellis-Bronnikov wormhole spacetime.
In principle, our results may be used to model any scenario (gravitational
or otherwise) where we have
a symmetric, single barrier potential in a wave equation.

\section*{\bf ACKNOWLEDGEMENTS}

\noindent The authors thank the anonymous  referee for very useful comments which helped us in 
improving the quality and presentation of the work reported in this article. JD thanks the Department of Physics, IIT Kharagpur, India for providing him with the opportunity
to work on this project during his tenure as a Masters student in IIT Kharagpur, India.
PDR thanks CTS, IIT Kharagpur, India for permitting her to use its facilities and  
IIT Kharagpur, India for support through a fellowship.


\begin{references}
\bibitem{lamb} H. Lamb, {\em On a Peculiarity of the wave‐system due to the free vibrations of a nucleus in an extended medium}, Proc. Lond. Math. Soc.,s1-32 (1) (1900), pp. 208-213, $10.1112/plms/s1-32.1.208$.
\bibitem{regge} T. Regge and J. A. Wheeler, {\em Stability of a Schwarzschild singularity}, Phys. Rev. {\bf 108}, 1063 (1957).
\bibitem{zerilli} F. J. Zerilli, {\em Effective potential for even-parity Regge-Wheeler gravitational perturbation equations}, Phys. Rev. Lett. {\bf 24}, 737 (1970).
\bibitem{Vishveshwara} C. V. Vishveshwara, {\em Scattering of gravitational radiation by a Schwarzschild black hole}, Nature {\bf 227}, 936 (1970).
\bibitem{Vishveshwara1} C. V. Vishveshwara, {\em Stability of the Schwarzschild metric}, Phys. Rev. D {\bf 1}, 2870 (1970).
\bibitem{Chandrasekhar} S. Chandrasekhar and S. Detweiler, {\em The quasi-normal modes of the Schwarzschild black hole}, Proc. R. Soc. Lond. {\bf A 344},441 (1975).
\bibitem{PT} G. P\"{o}schl and E. Teller, {\em Bemerkungen zur Quantenmechanik des anharmonischen Oszillators}, Z. Physik {\bf 83}, 143 (1933).
\bibitem{PT1} A. F. Cardona and C. Molina,{\em Quasinormal modes of generalized Pöschl-Teller potentials}, 	Class. Quant. Grav. {\bf 34} 245002 (2017).
\bibitem{visser} P. Boonserm and M. Visser, {\em Quasi-normal frequencies: key analytic results}, arXiv:1005.4483v3 [math-ph], JHEP, 1103:073 (2011).
\bibitem{Ligo1}  B. P. Abbott et. al, {\em Observation of gravitational waves from a binary black hole merger}, Phys. Rev. Letts. {\bf 116}, 061102 (2016).
\bibitem{aneesh} S. Aneesh, S. Bose and S. Kar, {\em Gravitational waves from quasinormal modes of a class of Lorentzian wormholes}, Phys. Rev. D {\bf 97}, 124004 (2018).
\bibitem{f(R)} S. Bhattacharyya and S. Shankaranarayanan, {\em Quasinormal modes as a distinguisher between general relativity and f(R) gravity: charged black holes},  arXiv:1803.07576v3 [gr-qc] (2018).
\bibitem{WKB} B. F. Schutz and C. M. Will, {\em Black hole normal modes - a semi-analytic approach}, Astrophys. J., {\bf L291}, 33 (1985).
\bibitem{ghatak} A. K. Ghatak, E. G. Sauter and I. C. Goyal, {\em Validity of the JWKB formula for a triangular potential barrier }, Eur. Jr. Phys., Vol.{\bf 18},
199 (1997).
\bibitem{mashhoon1} B. Mashhoon, {\em Quasi-normal modes of a black hole}, Proceedings of the third Marcel Grossmann meeting on General Relativity, edited by H. Ning (1983) pp. 599--608.
\bibitem{mashhoon2} V. Ferrari and B. Mashhoon, {\em Oscillations of a black hole}, Phys. Rev. Lett. {\bf 52}, 1361 (1984).
\bibitem{ferrari} V. Ferrari and B. Mashhoon, {\em New approach to the quasinormal modes of a black hole}, Phys. Rev. D {\bf 30}, 295 (1984).
\bibitem{blome} H.J. Blome and B. Mashhoon, {\em Quasi-normal oscillations of a Schwarzschild black hole}, Phys. Lett. A, Vol.{\bf 100 }, 231 (1984).
\bibitem{morris} M. Morris and K. S. Thorne, {\em Wormholes in spacetime and their use for interstellar travel: A tool for teaching general relativity}, Am. J. Phys., {\bf 56}, 395 (1988).
\bibitem{ellis} H. J. Ellis, {\em Ether flow through a drainhole: a particle model in general relativity}, J. Math. Phys. {\bf 14}, 104 (1973); K. A. Bronnikov, {\em Scalar tensor theory and scalar charge}, Acta Phys. Polon. {\bf B4}, 250 (1973)
\bibitem{cardoso} V.Cardoso, {\em Quasinormal modes and gravitational radiation in black hole spacetimes},  arXiv:0404093v1 [gr-qc], Section: {\bf 1.1.2} (2004). 
\bibitem{DLMF} Digital Library of Mathematical Functions: Section 9.2 (v); https://dlmf.nist.gov/9.2.
\bibitem{mathematica} Wolfram Research, Inc.,
{\em Mathematica, Version 10.0}, Champaign, IL (2014).
\bibitem{price} C. Gundlach, R. Price and J. Pullin, {\em Late-time  behavior of  stellar collapse and explosions.I. Linearized  perturbations}, Phys. Rev. {\bf D 49} ,883 (1994).
\bibitem{konoplya}   R.A. Konoplya and A. Zhidenko, {\em Quasinormal modes of black  holes: from astrophysics to string theory}, Rev. Mod. Phys. {\bf 83}, 793 (2011).
\bibitem{nollert} H. P. Nollert, {\em 
About the significance of quasi-normal Modes of Black Holes}, Phys. Rev. D {\bf 53},  4397-4402 (1996).
%\bibitem{ECO} V. Cardoso, E. Franzin and P. Pani, {\em Is the Gravitational-Wave Ringdown a Probe of the Event Horizon?}, Phys. Rev. Lett. {\bf 116}, 171101 (2016).
%\bibitem{gravastar}   C.Chirenti and L.Rezzolla,
%{\em Did GW150914 produce a rotating gravastar}, Phy. Rev. {\bf D 94}, 084016 (2016).
\bibitem{cardoso1} V. Cardoso, S. Hopper, C. F. B. Macedo, C. Palenzuela and Paolo Pani, {\em Echoes of ECOs: gravitational-wave signatures of exotic compact objects and of quantum corrections at the horizon scale}, Phys. Rev. D {\bf 94}, 084031 (2016).
\bibitem{krishnendu} N. V. Krishnendu, K. G. Arun, C. K. Mishra, {\em Testing the binary black hole nature of a compact binary coalescence},  	Phys. Rev. Lett. {\bf 119}, 091101 (2017).
\bibitem{Ligo} D. V. Martynov, E. D. Hall, B. P. Abbott et. al., {\em The sensitivity of the Advanced LIGO detectors at the beginning of gravitational wave astronomy},  arXiv:1604.00439v3 [astro-ph.IM] (2016). 
\bibitem{enhanced ligo} M. Page, J. Quin, J. L. Fontaine, C. Zhao, L. Ju and D. Blair, {\em Enhanced detection of high frequency gravitational waves using optically diluted optomechanical filters}, arXiv:1711.04469v2 [astro-ph.IM] (2017).
\bibitem{rectangular barrier} F. A. Handler and R. A. Matzner, {\em Gravitational wave scattering}, Phys. Rev. {\bf D 10}, Vol. 22 (1980).
\bibitem{kim} S. Kim, {\em Wormhole perturbation and its quasi-normal modes}, Prog. Theo. Phys. Suppl., No. {\bf 172}, 21 (2008).
\bibitem{konoplya1} R. A. Konoplya, A. Zhidenko, {\em Wormholes versus black holes: quasinormal ringing at early and late times}, JCAP {\bf 12}, 043 (2016).
\bibitem{kunz} J. L. Bl{\'a}zquez-Salcedo, X. Y. Chew and J. Kunz, {\em Scalar and axial quasinormal modes of massive static phantom wormholes}, Phys. Rev. D {\bf 98}, 044035 (2018).
\bibitem{kerr WH} P. Bueno, P. A. Cano, F. Goelen, T. Hertog and B. Vercnocke, {\em Echoes of Kerr-like wormholes}, Phys. Rev. D {\bf 97}, 024040 (2018).
\bibitem{paolo} P. Pani, {\em Advanced methods in black-hole perturbation theory},  arXiv:1305.6759v2 [gr-qc] (2013).
\bibitem{prony1}  L. London, J. Healy and  D. Shoemaker, {\em Modeling ringdown: beyond the fundamental quasinormal modes}, Phys. Rev. D {\bf 90}, 124032 (2014).
\bibitem{prony2} E. Berti, V. Cardoso, J. A. González and U. Sperhake, {\em Mining information from binary black hole mergers: a comparison of estimation methods for complex exponentials in noise}, Phys. Rev. D {\bf 75}, 124017 (2007).
\bibitem{prony3} O. P. F. Piedra, {\em Gravitino perturbations in Schwarzschild black holes},  	Int. J. Mod. Phys. D {\bf 20}:93--109 (2011).
\bibitem{CF} E.W. Leaver, {\em An analytic representation for the quasi-normal modes of Kerr black Holes}, Proc. R. Soc. Lond., \textbf{A 402}, 285 (1985). 



\end{references}
\end{document}